\documentclass[abbrv,aps,prl,twocolumn,showpacs,preprintnumbers,amsmath,amssymb,
superscriptaddress,nofootinbib]{revtex4}
\usepackage {graphicx}

\begin {document}

\title {Pressure-Driven Metal-Insulator Transition in Hematite from Dynamical
Mean-Field Theory}

\author {J.~Kune\v{s}}
\affiliation {Theoretical Physics III, Center for Electronic Correlations
and Magnetism, Institute of Physics, University of Augsburg, Augsburg
86135, Germany}

\author {Dm.~M.~Korotin}
\affiliation {Institute of Metal Physics, Russian Academy of Sciences,
620041 Yekaterinburg GSP-170, Russia}

\author {M.~A.~Korotin}
\affiliation {Institute of Metal Physics, Russian Academy of Sciences,
620041 Yekaterinburg GSP-170, Russia}

\author {V.~I.~Anisimov}
\affiliation {Institute of Metal Physics, Russian Academy of Sciences,
620041 Yekaterinburg GSP-170, Russia}

\author{P. Werner}
\affiliation{Theoretische Physik, ETH Zurich, 8093 Zurich, Switzerland}

\begin {abstract}
The Local Density Approximation combined with Dynamical Mean-Field Theory (LDA+DMFT method) is applied to the study of the paramagnetic and magnetically ordered phases of hematite Fe$_2$O$_3$ as a function of volume. 
As the volume is decreased, a simultaneous 1st order insulator-metal and high-spin to
low-spin transition occurs close to the experimental value of the critical volume. The high-spin insulating phase is destroyed
by a progressive reduction of the charge gap with increasing pressure, upon closing of which the high spin phase becomes unstable.
We conclude that the transition in Fe$_2$O$_3$ at $\approx$50 GPa can be described as an electronically driven volume collapse. 
\end {abstract}

% 74.25.Jb Electronic structure 
% 71.27.+a Strongly correlated electron systems; heavy fermions
% 71.30.+h Metal-insulator transitions and other electronic transitions
% 91.60.x Physical properties of rocks and minerals 
% 91.60.Gf High-pressure behavior

\maketitle

Metal-insulator transition (MIT) is one of the central topics in the physics
of strongly correlated electron materials. Despite considerable
progress in the theoretical understanding of this phenomenon in the past decade,
we are only starting to uncover the multitude of possible
transition scenarios in real materials. An important instance of the MIT
is the pressure driven transition accompanied by a change
of the local spin state (high spin (HS) to low spin (LS) transition)
seen in MnO \cite{yoo-05}, BiFeO$_3$ \cite{gav-08} or Fe$_2$O$_3$. 
The understanding of the pressure-driven HS-LS transition
and its relationship to eventual MIT and structural and/or volume changes
is relevant to a broader class of oxides -- often with important
geological implications -- such as FeO, CoO or Fe$_3$O$_4$.

In this Letter we study the spin and metal-insulator transition in
hematite ($\alpha$-Fe$_2$O$_3$) under pressure
using the numerical LDA+DMFT approach \cite{ldadmft}.
At ambient conditions, hematite is an antiferromagnetic insulator ($T_N$=956~K) with
the corundum structure \cite {Sgull-51}. The iron ions having a formal
Fe$^{3+}$ valency with five $d$ electrons give rise to a local HS state.
Photoemission studies \cite{Fujimori-86,lad-89} classified hematite as charge-transfer
insulator. A gap of 2.0-2.7~eV was inferred from the electrical conductivity data \cite{mochizuki}.
Under pressure, a first-order phase transition is observed at approximately 50~GPa 
(82\% of the equilibrium volume) at which the specific volume 
decreases by almost 10\% and the crystal-lattice symmetry is reduced (a
structure of the Rh$_2$O$_3$-II type is formed) \cite {Pasternak-99,Rozenberg-02,Liu-03}.
The high-pressure phase is characterized by a metallic conductivity
and absence of both the long-range magnetic order and the
HS local moment \cite {Pasternak-99}. Badro {\it et al.}  
showed that the structural transition actually precedes the electronic transition,
which is nevertheless accompanied by a sizable reduction of the bond lengths \cite{Badro-02}.

%COMPUTATIONAL THEORY *********
The investigations of the electronic structure of hematite 
using various theoretical approaches produced different
(and frequently controversial) results. Similar to many transition metal oxides, calculations within the local spin
density (LSDA) and generalized gradient (GGA) approximations
failed to reproduce the ground state characteristics 
under ambient conditions, underestimating values of the
magnetic moment of iron and of the energy gap \cite
{Sandratskii-96}. Within GGA, the structural
transition together with a transition into the metallic ferromagnetic
state with Fe ions in the LS state was predicted at a pressure of 14~GPa
\cite {Rollmann-04}, which is about one fourth of the
experimental value. 
The studies using the LSDA+U approach, which treats the local Coulomb
interaction on the static mean-field level \cite {Rollmann-04} %Punkkinen-99,
%Bandyopadhyay-04, Mazurenko-05
lead to improved values of the magnetic moment and the gap in the HS phase, but failed to capture the
metal-insulator transition. Like for some other oxides, LSDA+U improves the description
of the HS low-pressure phase, while LSDA appears more appropriate for the description of the LS high-pressure phase,
none of them covering the full pressure range. Thus, the DMFT method, which can capture
the evolution from strong to weak local correlations is a promising tool for
investigating the physics in the transition regime, as was shown for MnO \cite{mno-kunes}.
The only available DMFT study of Fe$_2$O$_3$ to-date was limited to the $t_{2g}$ band manifold finding 
a low-spin metallic phase at high pressure \cite {Kozhevnikov-07}.

In this work we use a Hamiltonian including explicitly all O-$p$ and Fe-$d$ orbitals, which
allows us to address the following questions. i) Is there an electronic transition without invoking a change in crystal structure?
ii) Is there a simultaneous metal-insulator and local moment transition
as found in the isoelectronic MnO \cite{mno-kunes}? iii) What is the nature
of the transition and how is it affected by temperature and long-range magnetic order? 
We use an implementation of LDA+DMFT, previously applied to NiO and described in Ref.~\onlinecite{kunes-nio}, which combines the first-principles
multi-band Hubbard Hamiltonian 
\begin{equation*}
%\label{eq:ham}
%\begin{split}
%\hat{H}=&\sum_{\mathbf{k}}H_{\alpha\beta}(\mathbf{k})\hat{a}_{\alpha}^{\dagger}(\mathbf{k})
%\hat{a}_{\beta}(\mathbf{k})-\varepsilon_{dc}(n_d)\hat{N}_d \\
%+&\sum_i U_{\alpha\beta}\hat{n}^d_{\alpha}(\mathbf{R}^{Fe}_i)\hat{n}^d_{\beta}(\mathbf{R}^{Fe}_i) %,
%\end{split}
\hat{H}=\sum_k\hat{\mathbf a}^{\dagger}_k{\mathbf H_k}
\hat{\mathbf a}_k-\varepsilon_{dc}(n_d)\hat{N}_d \\
+\sum_{i\in Fe} \hat{\mathbf n}^d_i{\mathbf U}^{dd}\hat{\mathbf n}^d_i
\end{equation*}
with DMFT equations.
Here, $\mathbf H_k$ is the LDA Hamiltonian matrix (38$\times$38 plus
spin degeneracy; two Fe$_2$O$_3$ units)
constructed on the basis of Fe-$d$ and O-$p$ Wannier orbitals represented
numerically on a uniform k-mesh in the first Brillouin zone \cite{DKorotin-08}. The
second term is the double-counting correction amounting to a constant
shift applied to Fe-$d$ site energies. The last term is the two-particle 
interaction at the Fe sites in the density-density approximation, parameterized with $U$=6.8~eV and $J$=0.86~eV  \cite{Anisimov}.
(Fitting the experimental photoemission spectra yields U of 8~eV according to Ref.~\onlinecite
{Fujimori-86} or  7~eV according to Ref.~\onlinecite {Bocquet-92}.)
For the double-counting correction $\varepsilon_{dc}$ we have adopted the
prescription $\varepsilon_{dc}=(N_{\text{orb}}-1)\bar{U}\bar{n}_d$ of Ref.~\onlinecite{kunes-nio},
but taking the self-consistent DMFT value for the average occupancy $\bar{n}_d$ rather
than its LDA value. 
%following the suggestion of Lichtenstein \cite{Licht}.
\begin{figure}
\includegraphics[height=\columnwidth,angle=270,clip]{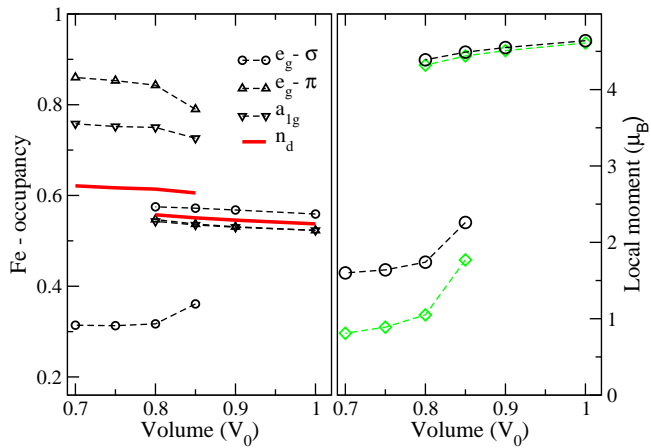}
\caption{\label{fig:1} Orbital occupancies (left panel)
at various specific volumes in PM state at T=580~K. The thick line
marks the average $d$ occupancy per orbital. Corresponding values of the screened (diamonds)
and instantaneous (circle) local moment
are shown in the right panel.}
\end{figure}
This approach, which can be viewed as a poor man's charge self-consistency loop, is necessary
for stabilization of the ambient pressure phase as discussed later.
For the solution of the $d$-shell impurity problem we have employed a
continuous time quantum Mote-Carlo (CT-QMC) solver \cite{ctqmc}. The calculations
were performed in the corundum $\alpha$-Fe$_2$O$_3$ structure with
rhombohedral (space-group R$\bar{3}$c) unit cell containing two formula units.
Throughout the paper the compressed phase is characterized by the volume
relative to the experimental ambient-pressure structure with lattice
parameter a$_0$=5.361~\AA, rhombohedral angle $\alpha$=55$^\circ$17$^\prime$,
and atomic coordinates Fe(0,0,0.355) and O(0.300,0,0.25).
Volume reduction is achieved by scaling the lattice constant.

In Fig.~\ref{fig:1} we show the evolution of the orbital occupancies and Fe local moment under compression 
in the absence of long-range magnetic order. The local
moment is measured as the square root of i) $\langle m_z^2 \rangle$ (instantaneous moment) or
ii) $\frac{1}{\beta}\int_0^{\beta}d\tau \langle m_z(\tau)m_z(0) \rangle$ (screened moment),
where $\beta$ is the inverse temperature. At the ambient conditions (volume $V_0$) we find
a HS solution with both definitions of local moment yielding essentially the value of 4.6~$\mu_B$.
Upon compression the HS phase survives down to 0.8$V_0$.
Below this volume only a LS phase with substantially reduced local moment is stable.
The difference between its screened and instantaneous values indicates that the LS moment
cannot be viewed a rigid object. The HS and LS solutions coexist in the range $0.8-0.85V_0$.
The Fe-$d$ orbital occupancies fit into the ionic picture of a 
fully spin-polarized $d$-shell in the HS state, and an orbitally polarized $d$-shell in 
the LS state. Nevertheless, the substantial $e_g^\sigma$ population in the LS phase reflects a
sizable covalent $p-d$ bonding (hybridization between O-$p$ and Fe-$e_g$ orbitals).

\begin{figure}
\includegraphics[height=\columnwidth,angle=270,clip]{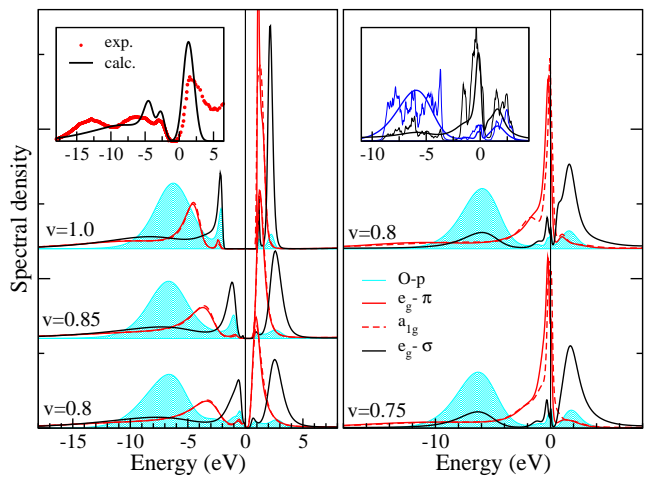}
\caption{\label{fig:2} (Color online) Evolution of the single-particle spectra with pressure
obtain in the PM state at  T=580~K. The HS solutions are shown in the left and LS solutions
in the right panel. All curves are normalized to one.
The left inset presents comparison of experimental PES \cite{lad-89}
and IPS \cite{ciccacci-91} data (points) to the calculated Lorentzian broadened $d$-spectral density
(solid line). The right inset shows a comparison of the v=0.8 DMFT (smooth lines) and non-interacting
LDA spectra (jagged lines): total $d$ spectral density (dark; black), total O-$p$ spectral
density (light; blue).}
\end{figure}
In Fig.~\ref{fig:2} we show the corresponding single-particle spectral densities.
Note the difference of the O-$p$ spectral weight above the chemical potential between
HS and LS solutions. In the LS state the $e_g^\sigma$-derived bands are above the chemical
potential, carrying some admixture of O-$p$ character with them, which leads a charge transfer
from the O-$p$ orbitals and an increase of $\bar{n}_d$ (see Fig. \ref{fig:1}). 
Relaxation of the electronic states, which compensates this charge redistribution
in real material, is mimicked by the self-consistency of the double counting term
in the present work. Comparison to the photoemission and
inverse photoemission data (see left inset of Fig. \ref{fig:2}) reveals a fair agreement
between the theory and experiment. The computed ambient pressure HS phase is found to be an insulator
with a charge gap of 2.6~eV. Upon compression the gap gradually shrinks
due to broadening of the bands and increasing of crystal-field splitting. 
At the high-pressure end of the HS stability region (0.8$V_0$) the gap is reduced 
to a fraction of an eV. The spectral density in the LS phase
closely resembles the non-interacting one (see right inset of Fig. \ref{fig:2}).
Unlike in the HS phase, there is some orbital polarization within the $t_{2g}(=a_{1g}$+$e_g^\pi)$ orbitals,
split due to distortion of FeO$_6$ octahedra.

\begin{figure}
\includegraphics[height=\columnwidth,angle=270,clip]{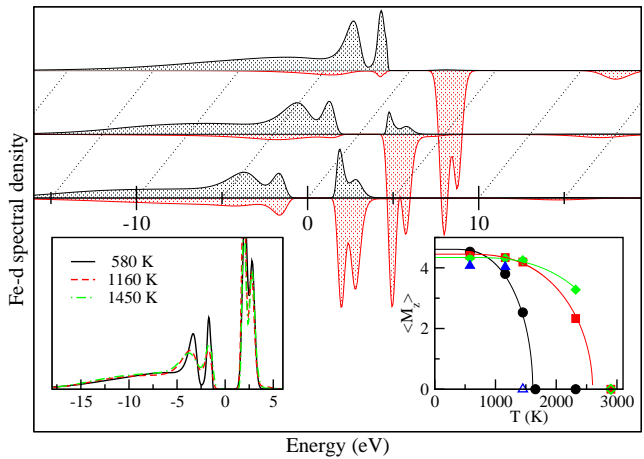}
\caption{\label{fig:3} Spin-polarized spectral Fe-$d$ spectral density at
ambient pressure for 580~K, 1160~K and 1450~K from the top to the bottom. The left
inset compares the same densities averaged over spin. The right inset shows
the staggered magnetization vs temperature curves for v=1.0 (circles; black),
v=0.9 (squares; red), v=0.85 (diamonds; green), and v=0.8 (triangles; blue).
At empty symbols only the LS solution exists. The lines depict mean-field fits.}
\end{figure}
Next, we address the role of magnetic long-range order. Unlike LDA or LDA+U,
for which static magnetic order is essential to capture a HS insulator,
in the present approach magnetic ordering can take place independently from the gap and local 
moment formation, as observed in many materials.
Using a symmetry that allows for the experimental AFM order we observe the following behavior.
At room temperature the HS solutions are magnetically ordered, while 
all LS solution remain paramagnetic.
In Fig. \ref{fig:3} we show the spin-polarized Fe-$d$ spectral densities at several
temperatures. Clearly,
the charge gap and the main spectral features are insensitive to the presence of AFM order.
The magnetization curves at reduced volume reflect the increase of the exchange coupling
with shortened atomic distances.

\begin{figure}
\includegraphics[height=\columnwidth,angle=270,clip]{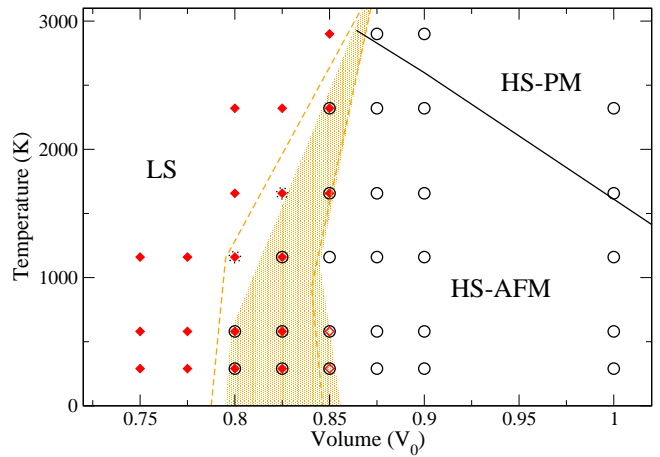}
\caption{\label{fig:4} Calculated V-T phase diagram of hematite. The symbols mark
the actual computation results: HS state (circle), LS state (diamond). Empty diamonds
mark LS solutions stable only if PM symmetry is constrained, dotted circles mark
HS solutions stable only if AFM order is allowed. The black line shows the N\'eel temperature
estimated from fits to v=1.0, 0.9 and 0.85 data. The shaded area marks the estimated
HS/LS coexistence region with PM constraint, the dashed lines mark the coexistence regime
when AFM order is allowed.}
\end{figure}

Our observations are summarized in the phase diagram shown in Fig.~\ref{fig:4}.
We distinguish three phases: i) paramagnetic LS, ii) antiferromagnetic HS and
iii) paramagnetic HS, separated by the line of N\'eel temperature $T_N$ and
the coexistence region of the HS and LS phases.
The evaluation of $T_N$ is affected by two approximations, both favoring the ordered phase.
First, the lack of non-local fluctuations inherent to DMFT and second, the approximate form of the 
local Coulomb interaction, limited to Ising terms, thus rendering the HS local state two- ($S_z=\pm5/2$) 
instead of sixfold degenerate ($S=5/2$). The latter approximation is also reflected
in the shape of the magnetization curves, which can be well fitted with a mean-field
theory based on the Brillouin function $B_{1/2}$ (see the inset of Fig.~\ref{fig:3}).
Given these approximations the calculated $T_N$ of 1600~K appears to be in reasonable agreement
with the experimental value of 956~K. 

The effect of AFM order on the HS-LS transition is to slightly shift the boundaries of the coexistence region
in favor of the HS phase. In particular, at the lowest
studied temperatures we find the LS solution at $0.85V_0$ to be unstable when AFM order is allowed.
Also we find the HS solutions at (0.8$V_0$, 1160~K) and (0.825$V_0$, 1660~K) stabilized by 
AFM order. Interestingly, in the vicinity of the HS/LS boundary we find 
the magnetic order with increasing temperature not to be destroyed by magnetic fluctuations, but via a collapse of the
constituting moments.

We propose two scenarios how the HS state can be destabilized and discuss them in the context of the
present results, the MnO results of Ref.~\onlinecite{mno-kunes} and the model calculations of Ref.~\onlinecite{werner-07}.
We start with a simple estimate of the charge gap considering the ionic energies
\begin{equation}
\label{eq:gap}
E_\text{g}\approx E(d^6)+E(d^4)-2E(d^5)-W=U+4J-2\Delta-W,
\end{equation}
where $2\Delta$ stands for the difference of $t_{2g}(=a_{1g}$+$e_g^\pi)$ and $e_g^\sigma$ energies (as in Ref. \onlinecite{werner-07}),
$W$ for the average of the $e_g$ and $t_{2g}$ bandwidths, $E(d^5)$ is the energy of the HS ionic groundstate
while $E(d^6)$ and $E(d^4)$ the ionic energies of the state with one more and one less electron, respectively. 
Similarly one can compare the energies of the ionic HS and LS states to get
\begin{equation}
\label{eq:lstate}
E(LS)-E(HS)=10J-4\Delta.
\end{equation}
The HS state becomes unstable when either (\ref{eq:gap}) or (\ref{eq:lstate}) becomes negative.
Thus depending on the parameters the HS state can be destroyed due to {\it gap closing} or
due to a {\it local state transition}. 
In charge-transfer compounds such as hematite $E_\text{g}$ is substantially reduced by the presence of the ligand bands. 
We found that a similar simple estimate for a 2-band model provides a fairly accurate description
of the phase boundaries found in Ref.~\onlinecite{werner-07}.

Comparing the present results with the earlier study of isoelectronic MnO \cite{mno-kunes} there is an obvious similarity in that 
the HS phase is an insulator while the LS phase is metallic.
However, there are two important differences in how the transition happens.
First, in MnO the transition is found to proceed through a continuous 
sequence of intermediate states while in hematite the transition is first order with a coexistence region 
of HS and LS solutions.
Second, we find the gap to almost close at the high-pressure end of the HS stability region in Fe$_2$O$_3$ while 
in MnO the transition starts with the appearance of an in-gap spectral density.
The latter observation leads us to suggest that the HS-LS/insulator-metal transition in hematite
proceeds through the {\it gap closing} mechanism while in MnO a {\it local state transition}
takes place as was argued in Ref.~\onlinecite{mno-kunes}. 
The gap closing picture of the transition in hematite is also supported by the shape of the HS/LS phase boundary,
showing that HS state can be destroyed by thermal fluctuations, contrary to the local state transition
picture where the HS state is expected to be favored at high temperature due to its higher entropy. 

Finally, we briefly address the reported structure and volume changes \cite{Pasternak-99,Rozenberg-02,Badro-02}. The
corundum and Rh$_2$O$_3$-II structures have similar local Fe environment and transitions between
the two were reported in other materials without significant changes of electronic structure. In this study we
do not attempt to distinguish between the two structural types. Finding 
the electronic HS-LS/insulator-metal transition in the high-volume phase shows that the electronic
transition is not implicated by the structural transition or volume collapse. Conversely the
electronic transition provides a natural explanation of the volume collapse as a consequence 
of emptying of the anti-bonding $e_g^\sigma$ bands in the LS phase and corresponding strengthening of the Fe-O bonds.

In summary, we have observed a simultaneous HS-LS/insulator-metal transition in the corundum phase of hematite 
close to the experimental volume in the numerical results obtained by LDA+DMFT. 
The long-range magnetic order is found to play a minor role in the transition.
We have discussed two possible scenarios, the {\it gap closing} and
the {\it local state transition}, and concluded that the transition in hematite, unlike in MnO, proceeds through the
gap closing mechanism. Our results are consistent with the picture of an electronically driven volume collapse.

J.K. acknowledges the support of SFB 484 of the Deutsche
Forschungsgemeinschaft. Support by the Russian Foundation for Basic
Research (project No. RFFI-07-02-00041) and Presidium of the Russian
Academy of Sciences (project No. 31, subprogram No. 3, Program No. 9) is
gratefully acknowledged.

\end {document}